\documentclass[aps,prb,twocolumn,10pt,superscriptaddress,longbibliography ]{revtex4-1}

\setcitestyle{numbers,square}

\usepackage{physics}
\usepackage[english]{babel}
\usepackage[utf8]{inputenc}
\usepackage{amssymb}
\usepackage{amsmath}
\usepackage[pdftex]{graphicx}
\usepackage{dsfont}
\usepackage{bm}
\usepackage[normalem]{ulem}
\usepackage{hyperref}
\hypersetup{
    colorlinks,%
    citecolor=blue,%
    linkcolor=blue,%
    urlcolor=blue
}

\providecommand{\ket}[1]{\vert #1\rangle} 
\providecommand{\bra}[1]{\langle #1\vert} 
\providecommand{\mean}[1]{\ensuremath{ \left\langle #1 \right\rangle} } 
\providecommand{\diff}[3]{\ensuremath{ \dfrac{\partial^{#1} {#3} }{\partial{#2}^{#1}}}} 

\usepackage{color}

\begin{document}
\title{Zitterbewegung and bulk-edge Landau-Zener tunneling in topological insulators}
\author{Gerson J.~Ferreira}
\author{Renan P. Maciel}
\affiliation{Instituto de F\'isica, Universidade Federal de Uberl\^andia, Uberl\^andia, Minas Gerais 38400-902, Brazil}

\author{Poliana H. Penteado}
\affiliation{International Institute of Physics, Universidade Federal do Rio Grande do Norte, 59078-970 Natal-RN, Brazil}
\affiliation{Instituto de F\'isica de S\~ao Carlos, Universidade de S\~ao Paulo, S\~ao Carlos, S\~ao Paulo 13560-970, Brazil}

\author{J. Carlos Egues}
\affiliation{Instituto de F\'isica de S\~ao Carlos, Universidade de S\~ao Paulo, S\~ao Carlos, S\~ao Paulo 13560-970, Brazil}

\date{\today}
\begin{abstract}
We investigate the ballistic zitterbewegung dynamics and the Landau-Zener tunneling between edge and bulk states of wave packets in two-dimensional topological insulators. In bulk, we use the Ehrenfest theorem to show that an external in-plane electric field not only drifts the packet longitudinally, but also induces a transverse finite side-jump for both trivial and topological regimes. For finite ribbons of width $W$, we show that the Landau-Zener tunneling between bulk and edge states vanishes for large $W$ as their electric field-induced coupling decays with $W^{-3/2}$. This is demonstrated by expanding the time-dependent Schrödinger equation in terms of Houston states. Hence we cannot picture the quantum spin Hall states as arising from the zitterbewegung bulk trajectories `leaking' into the edge states, as proposed in Phys. Rev. B \textbf{87}, 161115 (2013).
\end{abstract}
\maketitle

\section{Introduction}

The dynamics of wave packets in multiband systems present a variety of interesting physical phenomena, e.g., the early studies of the Landau-Zener tunneling (LZT)  \cite{landau1932zener1, landau1932zener2, Zener1932, Stueckelberg1932, majorana1932zener}, the ballistic spin resonance in multichannel spin-orbit coupled quantum wires \cite{frolov2009ballistic, hachiya2014}, and the dynamics of unusual spin textures in spin-orbit coupled two-dimensional electron gases \cite{Schliemann2003SFET, Bernevig2006PSH, Koralek2009PSH, Walser2012PSH, Schliemann2016ControlPSH, Schliemann2017PSH}. Moreover, the trembling motion, or zitterbewegung, is a particularly interesting dynamics that arises due to the spin-orbit coupling in quantum wells \cite{Schliemann2005, Schliemann2006, Schliemann2007SideJump, Esmerindo2007zitter, Ho2014Persistentzitter} and topological insulators (TIs) \cite{Shi2013zitter, Kolodrubetz2016Nonadiabatic, Yar2017zitterTIEf}.

The topological insulators \cite{Bernevig2006BHZ, Molenkamp2007QSHE, Kane2010Review, Zhang2011ReviewTI, *[{For an all-electron 2D topological insulator in ordinary III-V quantum wells see: }] [{.}] Erlingsson2015}, are characterized by band inversions that lead to symmetry-protected helical edge/surface states with Dirac-like dispersion within the bulk gap. Recently, it was proposed  \cite{Shi2013zitter} that the semiclassical zitterbewegung trajectories of electrically-driven carriers in two-dimensional (2D) TIs could provide an intuitive picture for the dynamical emergence of the edge states. On the other hand, it is well known that in GaAs quantum wells the zitterbewegung is accompanied by a finite ballistic side-jump  \cite{Schliemann2007SideJump}. This process leads to spin polarization at the edges, constituting a ballistic spin Hall effect in narrow wires \cite{Xu2006zitterSHE}. In the diffusive regime, the side-jump accumulated after successive Markovian scatterings was recently related to the Rashba-Edelstein effect \cite{Vignale2016Edelstein}.

The zitterbewegung is usually calculated in bulk, while the edge states exist only at the borders of finite size samples. Therefore, it is interesting to investigate the effects of both the edge states and the quantum confinement on the ballistic dynamics of wave packets in TI ribbons in the presence of electric fields. More importantly, does the corresponding zitterbewegung trajectories indeed bear any connection with the helical edge states as proposed in Ref.~\onlinecite{Shi2013zitter}?

\begin{figure}[ht!]
  \centering
  \includegraphics[width=\columnwidth,keepaspectratio=true]{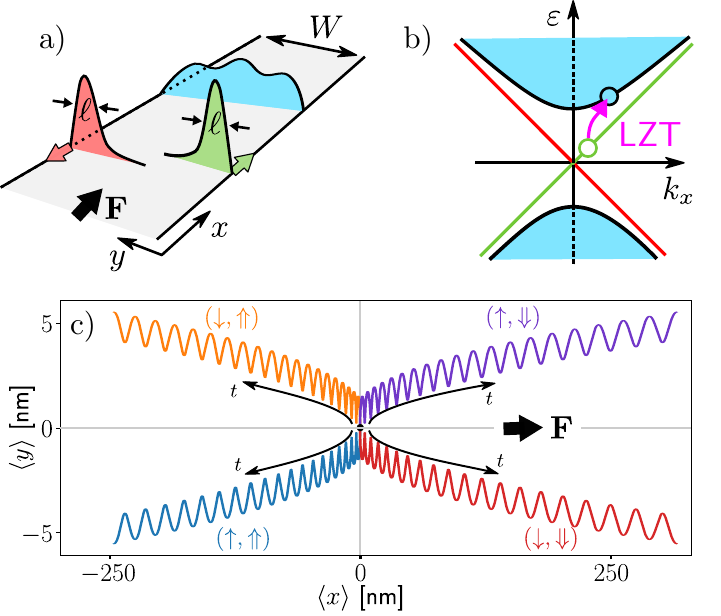}
  \caption{(a) Illustration of a BHZ ribbon of width $W$ with edge states, characterized by a penetration length $\ell$, and a bulk state (blue). (b) Sketch of the BHZ band structure in the nontrivial regime with linear edge state branches and bulk bands. The arrow indicates the Landau-Zener tunneling between an edge state and a bulk band promoted by the electric field $\bm{F}$. (c) zitterbewegung trajectories $[\mean{x}(t), \mean{y}(t)]$ in bulk for $0 < t < 5$~ps for different initial spin $s = \{\uparrow, \downarrow\}$ and pseudospin $\sigma_z = \{\Uparrow, \Downarrow\}$. For larger $t\gtrsim 50$~ps the motion along $y$ saturates into a finite side-jump $|y(\infty)| \approx 18.75$~nm.}
  \label{fig1}
\end{figure}

In this paper, we address this issue by investigating the ballistic dynamics of electrically-driven wave packets in 2D TIs fully accounting for the LZT between bulk and edge states. We model our system by the Bernevig-Hughes-Zhang (BHZ) Hamiltonian  \cite{Bernevig2006BHZ} in the presence of an external electric field $\bm{F}$. For a ribbon of width $W$, as illustrated in Figs.~\ref{fig1}(a) and \ref{fig1}(b), we show that the LZT is characterized by the nondiagonal terms of the Berry connection matrix $\mathcal{A}_{k_x}^{n,n'}$. For $W \gg \ell$, where $\ell$ is the typical length of the edge states, we find that the LZT vanishes with $W^{-3/2}$. In addition, we analyze the semiclassical zitterbewegung trajectories in bulk, which always show a finite ballistic side-jump  \cite{Schliemann2007SideJump} towards a direction that strongly depends on the initial conditions, Fig.~\ref{fig1}(c).

For a finite ribbon, we solve the Schr\"odinger equation numerically and study the full dynamics of a wave packet. Consistently with $\mathcal{A}_{k_x}^{n,n'}$ vanishing for $W \gg \ell$, we show that a Gaussian wave packet bounces off the borders of the ribbon unaffected by the presence of the edge states. These results do not support the idea from Ref.~\onlinecite{Shi2013zitter} that the helical edge states emerge from the bulk zitterbewegung trajectories, which would provide a semi-classical picture of the quantum spin Hall effect (QSHE). Instead, the zitterbewegung and finite ballistic side-jump seen here are compatible with the GaAs Rashba-Edelstein effect from Ref.~\onlinecite{Vignale2016Edelstein}, which was already suggested \cite{Schliemann2007SideJump} as a semi-classical picture of the spin Hall effect (SHE).

This paper is organized as follows. In Sec.~\ref{sec:model}, we present the model Hamiltonian and the numerical parameters. In Sec.~\ref{sec:LZT}, we discuss the Landau-Zener tunneling and its dependence on the system size. In Secs.~\ref{sec:zitter} and \ref{sec:wavepacket}, we present results for the zitterbewegung, using the Ehrenfest theorem, and the full time evolution of a wave-packet for large systems, respectively. We finally summarize our findings in Sec.~\ref{sec:conclusion}.

\section{Model system}
\label{sec:model}

We consider the BHZ Hamiltonian  \cite{Bernevig2006BHZ}

\begin{equation}
 H = C - D \bm{k}^2 + s A k_x \sigma_x + A k_y \sigma_y +(M-B\bm{k}^2)\sigma_z,
 \label{eq:BHZ}
\end{equation}
where $s=\pm 1$ labels the spin-up ($\uparrow$) and -down ($\downarrow$) subspaces, $\bm{\sigma}$ are the Pauli matrices  acting on the pseudospin subspace $\{E_1, H_1\} = \{\Uparrow, \Downarrow\}$ of the confined electron and hole states of the quantum well, and $\bm{k} = (k_x, k_y)$ are the in-plane momenta. Unless otherwise specified, we choose typical values for the parameters~ \cite{Molenkamp2007QSHE}: $C=6.5$~meV, $A = 375$~meV~nm, $B=-1120$~meV~nm$^2$, $D=-730$~meV~nm$^2$, and a negative Dirac mass $M = -10$~meV, which gives rise to  helical edge states as illustrated in Figs.~\ref{fig1}(a) and \ref{fig1}(b). An electric field $F \sim 10^{-3}$~mV/nm along $\hat{x}$ is introduced by the time-dependent vector potential $e\bm{A}(t) = -eFt\hat{x}$ via minimal coupling, $\bm{k} \rightarrow \bm{k} - e\bm{A}(t)/\hbar$. This gauge preserves $(k_x, k_y)$ as good quantum numbers in bulk, but it makes $H \rightarrow H_t$ time-dependent. 

For finite ribbons, the confinement is introduced via a $y$-dependent mass potential  \cite{Michetti2012EdgeStates, Ferreira2013Magnetic} $M \rightarrow M(y)$ given by

\begin{equation}
 M(y) = M_i + (M_o-M_i)\Big[1 \pm \dfrac{1}{2} \tanh \Big(\dfrac{y \pm W/2}{\gamma}\Big) \Big],
  \label{eq:mass}
\end{equation}
where $M_i$ is the mass gap of the system, $M_o$ is the mass of the confining barriers, and $\gamma$ specifies whether the potential profile is sharp ($\gamma \rightarrow 0$) or smooth. Unless otherwise specified, we consider the hard-wall limit, which corresponds to $\gamma \rightarrow 0$ and $M_o \rightarrow \infty$. Due to the confinement, $k_y$ is now quantized.

\begin{figure}[ht!]
  \centering
  \includegraphics[width=\columnwidth,keepaspectratio=true]{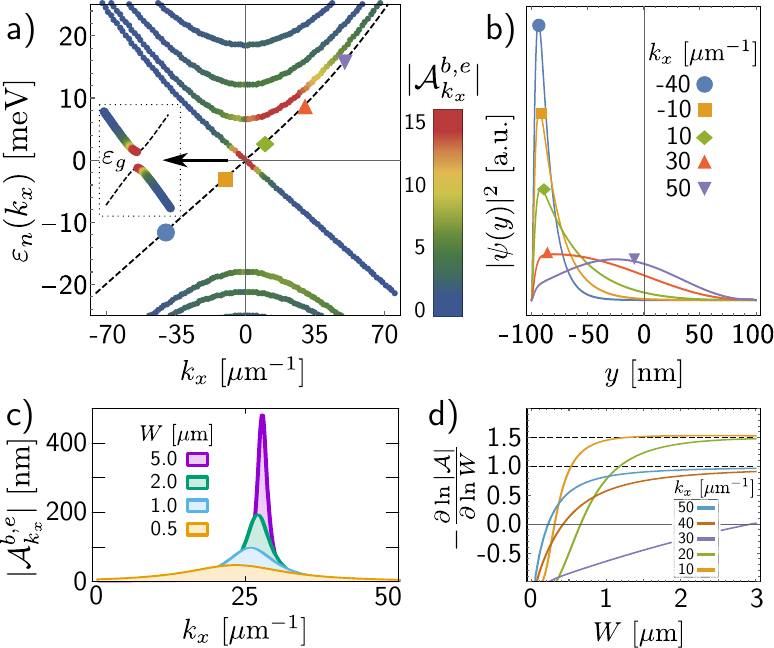}
  \caption{
  (a) Band structure $\varepsilon_n(k_x)$ of a $W=200$~nm ribbon. The color code in each band represents its Berry connection coupling $|\mathcal{A}_{k_x}^{b,e}|$ (in nm) to the selected edge branch (dashed). The inset at $k_x \approx 0$ shows the small hybridization gap $\varepsilon_g = 0.4$~meV.
  (b) Probability density of the edge state for different $k_x$'s, indicated by the symbols. As $k_x$ increases, the state becomes extended (bulk-like).
  (c) $|\mathcal{A}_{k_x}^{b,e}|$ between the first bulk quantized conduction band and the edge branch. As $W$ increases, the peak becomes sharp at $k_x = k_c \approx 28$~$\mu$m$^{-1}$.
  (d) Logarithmic derivative of $\ln|\mathcal{A}_{k_x}^{b,e}|$ with respect to $W$. For large $W \gg \ell$ ($\ell \approx |A/M| = 37.5$~nm), $|\mathcal{A}_{k_x}^{b,e}| \propto W^{-3/2}$ for $k_x < k_c$ and $|\mathcal{A}_{k_x}^{b,e}| \propto W^{-1}$ for $k_x > k_c$.
  }
  \label{fig2}
\end{figure}

\section{Landau-Zener tunneling}
\label{sec:LZT}

The original Landau-Zener formula  \cite{landau1932zener1, landau1932zener2, Zener1932, Stueckelberg1932, majorana1932zener}, developed for a pair of linear bands anticrossing, can be directly applied to the pair of hybridized edge-state branches arising in the ribbon geometry [see the gap $\varepsilon_g$ in the inset of Fig.~\ref{fig2}(a)]. Therefore the tunneling probability between edge states is given by

\begin{equation}
 \gamma_{ee} \approx \exp\Big[-\dfrac{2\pi\varepsilon_g^2}{\hbar v_f \; eF}\Big],
 \label{eq:zeneree}
\end{equation}
where $v_f\approx A/\hbar$ is the Fermi velocity of the nearly linear-in-$k_x$ edge branches. Next we analyze the LZT between edge and bulk states.

Consider the time-dependent Hamiltonian $H_t$ in the presence of an electric field as discussed above. Since $[H_t, k_x(t)] = 0$, we can write the solution as $\Psi(x,y,t) = e^{i k_x(t) x} \psi_{k_x(t)}(y,t)$. The unknown term $\psi_{k_x(t)}(y,t)$ can be expanded into Houston functions   \cite{marder2010condensed} $\varphi_{k_x(t),n}(y)$, which are solutions of an instantaneous Schr\"odinger equation $H_t  \varphi_{k_x(t),n}(y) = \varepsilon_n\big[k_x(t)\big]\varphi_{k_x(t),n}(y)$, with $n$ labeling both the quantized bulk bands ($n=b$) and the helical edge states ($n=e$) shown in Fig.~\ref{fig2}. Since $t$ is taken as a simple parameter in this auxiliary equation, $\varepsilon_n(k_x)$ is the band structure of the BHZ ribbon in the absence of an electric field. Ultimately the Houston functions provide a complete time-dependent basis set, such that the full expansion reads

\begin{equation}
 \psi_{k_x(t)}(y,t) = \sum_n \alpha_n(t) \varphi_{k_x(t),n}(y),
 \label{eq:psi}
\end{equation}
in which $\alpha_n(t)$ are time-dependent coefficients. Applying this expansion to the time-dependent Schr\"odinger equation we obtain

\begin{equation}
 \left[i\hbar \diff{}{t}{} - \varepsilon_n\big(k_x(t)\big)\right]\alpha_{n}(t) = -eF \sum_{n'} \mathcal{A}_{k_x(t)}^{n,n'} \alpha_{n'}(t),
 \label{eq:HoustonC}
\end{equation}
where $\mathcal{A}_{k_x(t)}^{n,n'} = i \bra{k_x, n} \frac{\partial}{\partial k_x}\ket{k_x, n'} |_{k_x \rightarrow k_x(t)}$ are the $(n, n')$ elements of the Berry connection matrix, and $\ket{k_x, n}$ is the Dirac ket for the Houston function $\varphi_{k_x,n}(y)$. For $n \neq n'$ ($\varepsilon_n \neq \varepsilon_{n'}$), $\mathcal{A}_{k_x(t)}^{n,n'}$ can be written as

\begin{equation}
 \mathcal{A}_{k_x(t)}^{n,n'} = \dfrac{i}{\varepsilon_{n'} - \varepsilon_{n}} \bra{k_x,n} \dfrac{\partial H}{\partial k_x} \ket{k_x,n'}\Big|_{k_x \rightarrow k_x(t)},
 \label{eq:BerryNN}
\end{equation}
which is most relevant near band anticrossings as $\varepsilon_n\big(k_x(t)\big) \approx \varepsilon_{n'}\big(k_x(t)\big)$. These nondiagonal terms connect different eigenstates $(n, n')$ via the electric field $eF$, i.e.,~it leads to the Landau-Zener tunneling. If restricted to a single band $n$, the usual diagonal Berry connection $\mathcal{A}_{k_x(t)}^{n,n}$ and Eq.~\eqref{eq:HoustonC} describe the adiabatic time-evolution, while the complete expression for many bands fully describes the quantum unitary evolution.

In general, the LZT probabilities are obtained by solving Eq.~\eqref{eq:HoustonC} using the steepest descent approximation~ \cite{marder2010condensed} around the extremum of the energy difference $\Delta \varepsilon_{nn'} = \varepsilon_n\big(k_x(t)\big) - \varepsilon_{n'}\big(k_x(t)\big)$, assuming that $\mathcal{A}_{k_x(t)}^{n,n'}$ varies slowly so it can be approximated by its value at $\Delta\varepsilon_{nn'}=0$. Here we cannot apply this method since bulk- and edge-state dispersions approach each other asymptotically as $k_x$ increases. Instead, we focus on the properties of the couplings $\mathcal{A}_{k_x(t)}^{n,n'}$ shown in Figs.~\ref{fig2} (a),~\ref{fig2}(c), and~\ref{fig2} (d).

Interestingly, for $W \gg \ell$, in which $\ell$ is the localization length of the edge states, the LZT between edge and bulk bands vanishes. This can be seen already from the normalization of these states. For an extended bulk state, the normalization goes as $\propto 1/\sqrt{W}$. On the other hand, the edge states ($\ket{k_x, e} \propto e^{-\tilde{y}/\ell}$, for an edge at $\tilde{y} = y-W/2 \approx 0$) are localized within a length scale $\ell \equiv \ell(k_x)$ from the edges [see Figs.~\ref{fig1}(a) and \ref{fig2}(b)]; at $k_x=0$, $\ell(0) \approx |A/M|$  \cite{Michetti2012EdgeStates}. In this case, the edge state normalization does not depend on the ribbon width. Additionally, the overlap between bulk and edge state wave functions is only finite near the edge, where the bulk state is qualitatively $\ket{k_x, b} \propto \frac{1}{\sqrt{W}}\sin(\pi \tilde{y}/W) \approx \pi \tilde{y}/W^{3/2}$, therefore the coupling $\mathcal{A}_{k_x(t)}^{b,e} \propto W^{-3/2}$; this is valid for small $k_x$. As $k_x$ increases, the edge states become extended, see Figs.~\ref{fig2}(a) and \ref{fig2}(b). Indeed for a semi-infinite system \cite{Volkov1, Volkov2}, $k_y$ switches from purely imaginary (evanescent wave with $\ell = \Im\{k_y\}^{-1}$) to purely real (bulk-like, oscillatory wave) at the $k_x = k_c$ point, where the linear edge branch enters the bulk band. Consequently for $k_x > k_c$ the normalization of the edge branch becomes bulk-like, i.e.,~$\propto 1/\sqrt{W}$, and the coupling scales as $\mathcal{A}_{k_x(t)}^{b,e} \propto W^{-1}$. Both scalings yield a vanishing $\mathcal{A}_{k_x(t)}^{b,e}$ for large $W$. 

The numerical evaluation shown in Fig.~\ref{fig2} confirms the asymptotic scalings of  $\mathcal{A}_{k_x(t)}^{b,e}$. In Fig.~\ref{fig2}(a) we calculate the coupling from a reference edge branch (dashed line) to all other edge and confined ribbon bands. The coupling intensity is indicated by the color code. The coupling $|\mathcal{A}^{e,e'}_{k_x(t)}|$ between the edge states is approximately a Lorentzian peak with broadening $\sim |2M/A|$, and intensity $\sim|A/\varepsilon_g|$ at $k_x=0$. For large $W\rightarrow \infty$ this coupling diverges as the gap closes, $\varepsilon_g =\varepsilon_n - \varepsilon_{n'} \rightarrow 0$. Similarly, in Figs.~\ref{fig2}(a) and \ref{fig2}(c), the coupling between the dashed edge branch and the first bulk band shows a sharp peak for $W\rightarrow\infty$ at $k_x = k_c$, which matches the point where the edge branch transitions from localized to extended [Fig.~\ref{fig2}(b)].

In Fig.~\ref{fig2}(d) we show the logarithmic derivative of the coupling, which for $\mathcal{A}_{k_x(t)}^{b,e} \propto W^{-p}$ yields $-\frac{\partial \ln \mathcal{A}}{\partial \ln W} = p$. 
For $k_x < k_c$, all lines approach $p=3/2$ asymptotically, while for $k_x > k_c$ they approach $p=1$. These scalings show that already for $W \approx 1$~$\mu$m the coupling between bulk and edge  states is negligible. Effectively, there is no LZT between edge and bulk states for large samples. 

\section{zitterbewegung}
\label{sec:zitter}

The zitterbewegung is the oscillatory motion of the mean value of the coordinates $[\mean{x}(t)$, $\mean{y}(t)]$, which is usually calculated in the broad wave-packet (plane-wave) limit, $\psi(t) \propto \exp[i \bm{k}(t)\cdot\bm{r}]$. To gain further insight about the ballistic dynamics of wave packets in TIs, we calculate these trajectories using the Ehrenfest theorem \cite{prezhdo2000quantized}. Omitting the time argument for brevity, i.e.,~$\mean{\mathcal{O}} \equiv \mean{\mathcal{O}}(t) = \bra{\psi(t)}\mathcal{O}\ket{\psi(t)}$, the coupled set of equations of motion reads 

\begin{align}
\dfrac{d\mean{x}}{dt} &= -\dfrac{2}{\hbar}\Big(D+B\mean{\sigma_z}\Big) k_x(t) + \dfrac{As}{\hbar} \mean{\sigma_x},
\label{eq:xt}\\
\dfrac{d\mean{y}}{dt} &= -\dfrac{2}{\hbar}\Big(D+B\mean{\sigma_z}\Big) k_y^0  + \dfrac{A}{\hbar} \mean{\sigma_y},
\label{eq:yt}\\
\dfrac{d{\mean{\bm{\sigma}}}}{dt} &= \dfrac{2}{\hbar} \bm{\Omega} \times \mean{\bm{\sigma}},
\label{eq:sigma}
\end{align}
where the pseudospin $\mean{\bm{\sigma}}(t)$ precession is given by the vector $\bm{\Omega}(t) = \{ sA k_x(t), Ak_y^0, M-B|\bm{k}(t)|^2\}$, with $k_x(t) = k_x^0 + eF t/\hbar$, and $(k_x^0, k_y^0)$ are the initial momentum. The initial pseudospin $\mean{\bm{\sigma}}(0) = \bm{\sigma}^0$ is arbitrary. For simplicity, we take $x(0) = y(0) = 0$. The equations above show that the dynamics of $\mean{x}(t)$ and $\mean{y}(t)$ have contributions from $\mean{\sigma_x}(t)$ and $\mean{\sigma_y}(t)$. These oscillate due to the pseudospin precession given by Eq.~\eqref{eq:sigma}, which gives rise to the zitterbewegung. Note that even for $k_y^0 = 0$, a finite transverse dynamics is set by the $\mean{\sigma_y}(t)$ contribution to $\mean{y}(t)$.

In the presence of the electric field, all possible initial conditions lead to a finite ballistic side-jump. This can be seen from the equations of motion noticing that, as time flows, $k_x(t)$ grows and the pseudospin precession is dominated by the $z$ component of $\bm{\Omega}(t) \approx \{0, 0, -B k_x^2(t)\}$. Therefore $\mean{\sigma_z}(t)$ asymptotically approaches a constant $\sigma_z^\infty$, while $\mean{\sigma_x}(t)$ and $\mean{\sigma_y}(t)$ precess around a zero average with increasing frequency $2\Omega_z(t)/\hbar$. The final value for $\sigma_z^\infty$ strongly depends on the initial conditions, ultimately affecting the asymptotic velocities in Eqs.~\eqref{eq:xt} and \eqref{eq:yt}. More importantly, the contributions from the last terms of these equations vanish on a time-average, thus ceasing the zitterbewegung and the transverse dynamics, resulting in a finite side-jump $y(\infty)$. 

\begin{figure}[ht!]
  \centering
  \includegraphics[width=\columnwidth,keepaspectratio=true]{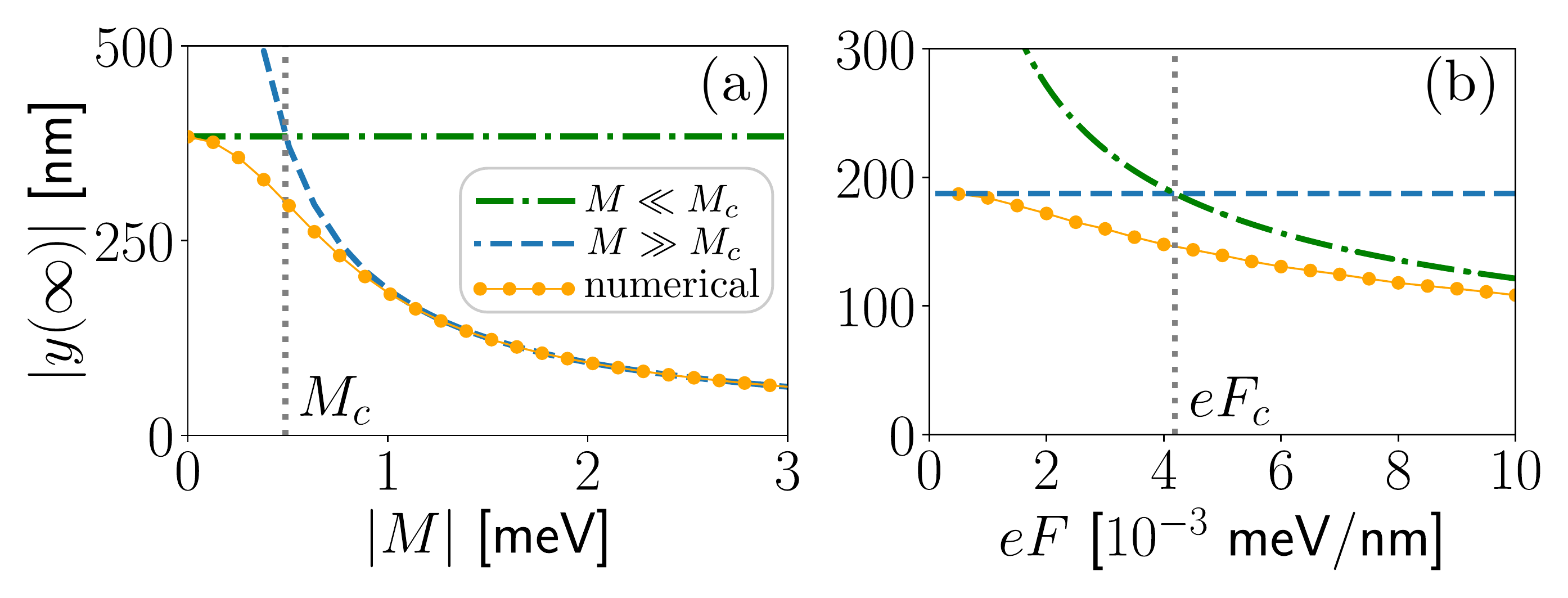}
  \caption{
  The asymptotic limit $|y(\infty)|$ of the finite ballistic side-jump as a function of (a) $M$ and (b) $eF$. For large $M \gg M_c$ (or $eF \ll eF_c$) the LZT is suppressed and the side-jump is independent of $eF$. In panel (a) we use $eF = 10^{-3}$~meV/nm, and in (b) $|M| = 1$~meV. In both cases $\bm{k}^0 = 0$, $s=\uparrow$, and $\sigma_z^0 = \Uparrow$. The dashed lines are the limiting cases from Eq.~\eqref{eq:yinf}, while the vertical dotted line marks the critical mass $M_c$ or electric field $eF_c$. The circles (orange) are the numerical data, which transits between both limiting cases.}
  \label{fig3}
\end{figure}

To express $y(\infty)$, let us analyze Eqs.~\eqref{eq:xt}-\eqref{eq:sigma} in the limits of small and large $M$. First, recall that the Landau-Zener tunneling is characterized by the Berry connection. In bulk, this coupling intensity is $|eF\mathcal{A}| \propto |eFA/M|$ at $\bm{k} = 0$, which becomes relevant if $|eF\mathcal{A}| \gg |2M|$ (band gap). For large and small $M$, we find 

\begin{equation}
    |y(\infty)| \approx
    \begin{cases}
        \sqrt{\dfrac{A\pi}{8eF}}, &\text{ for } |M| \ll M_c,
    \\ \\
        \dfrac{A}{2|M|}, &\text{ for } |M| \gg M_c.
    \end{cases}
    \label{eq:yinf}
\end{equation}
The transition point is defined by a critical mass $M_c \equiv \sqrt{2AeF/\pi}$, or critical field $eF_c \equiv \pi M^2/2A$. The solution for $M \ll M_c$ can be obtained analytically from Eqs.~\eqref{eq:xt}-\eqref{eq:sigma}, while for $M \gg M_c$ the expression above was extracted from fitting the numerical data. These expressions are compared to the numerical data in Fig.~\ref{fig3} with great accuracy.

The Ehrenfest dynamics for $M = -10$~meV, different initial spin $s=(\uparrow, \downarrow)$, and pseudospin $\sigma_z^0 = (\Uparrow, \Downarrow)$ are shown in Fig.~\ref{fig1}(c) for $0 < t < 5$~ps. The initial momentum is $(k_x^0, k_y^0) = (10^{-3}, 0)$~nm$^{-1}$ in all cases. The trajectories show the oscillatory behavior (zitterbewegung) due to the pseudospin precession. Note that since $k_y^0 = 0$, all the dynamics along the transverse direction ($y$) are induced by $\mean{\sigma_y}(t)$ in Eq.~\eqref{eq:yt}. As discussed above, for large $t$, the trajectories bend inwards as the transverse motion saturates. The zitterbewegung vanishes for $t \gtrsim 50$~ps with a finite side-jump of $|y(\infty)| \approx A/2|M| = 18.75$~nm.

\begin{figure}[ht!]
  \centering
  \includegraphics[width=\columnwidth,keepaspectratio=true]{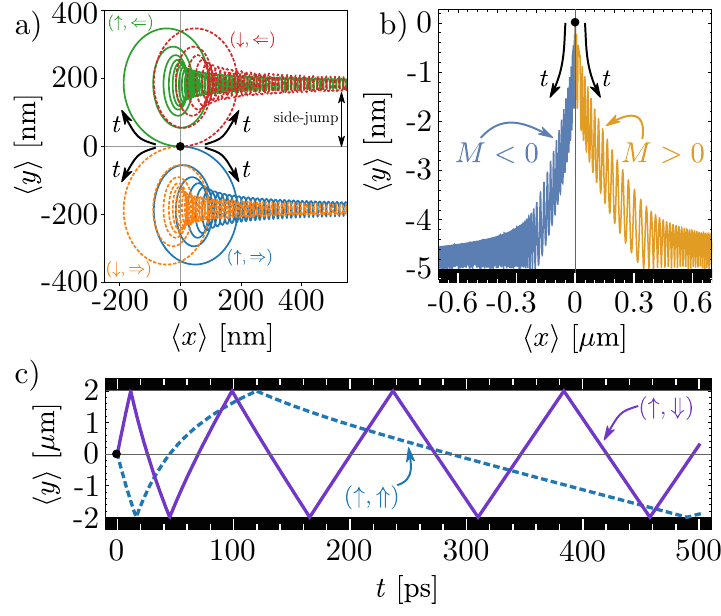}
  \caption{
  (a) Bulk zitterbewegung trajectories for $M = -1$~meV, $\bm{k}^0 = 0$, and different initial spins $s = \{\uparrow, \downarrow\}$, and pseudospins $\sigma_x^0 = \{\Rightarrow, \Leftarrow\}$, as indicated. The reduced $M$ leads to a larger side-jump $\sim \pm 200$~nm.
  (b) Incorrect skipping orbits at the edge ($y=-5$~nm) appears for both trivial ($M>0$) and nontrivial ($M<0$) topological regimes if the Ehrenfest equations Eqs.~\eqref{eq:xt}-\eqref{eq:sigma} are used beyond their limit of validity. Here $(k_x^0, k_y^0) = (10^{-3}, 0)$~nm$^{-1}$, initial $(s, \sigma_z^0) = (\uparrow, \Uparrow)$, and $M=\pm 10$~meV.
  (c) Qualitative zitterbewegung $\mean{y}$ vs $t$ on a large $W = 4$~$\mu$m ribbon for different initial conditions $s = \uparrow$ and $\sigma_z^0 = \{\Uparrow, \Downarrow\}$. In all cases $(k_x^0, k_y^0) = (10^{-3}, 10^{-2})$~nm$^{-1}$, $M = -10$~meV. At the edges $y = \pm W/2$ the trajectory bounces with specular reflection as $k_y \rightarrow -k_y$ and $\mean{\sigma_y} \rightarrow -\mean{\sigma_y}$ in Eq.~\eqref{eq:yt}.
  }
  \label{fig4}
\end{figure}

The initial pseudospin $\bm{\sigma}^0$ also affects significantly the dynamics. In Fig.~\ref{fig4}(a) we consider $\bm{\sigma}^0$ as the eigenstates of $\sigma_x$, which we label as $\sigma_x^0 = \{\Rightarrow, \Leftarrow\}$. Since at $\bm{k} \approx 0$ $\bm{\Omega}(t) \parallel \hat{z}$, the precession is approximately a circular motion of $\mean{\sigma_x}(t)$ and $\mean{\sigma_y}(t)$ on the $\mean{\sigma_z} = 0$ plane of the pseudospin Bloch sphere. This yields the nearly circular motion seen in Fig.~\ref{fig4}(a) for $M=-1$~meV, which is distorted by the electric field $F = 10^{-3}$~mV/nm. As time flows, the circular motion ceases and converges towards a side-jump of $\sim \pm 200$~nm. For $M = -10$~meV the dynamics is qualitatively equivalent to Fig.~\ref{fig4}(a), but with a much faster precession that becomes difficult to visualize. Figures \ref{fig1}(c) and \ref{fig4}(a) were obtained for $M < 0$. However, nearly identical results are obtained in the trivial regime ($M>0$) simply by mirroring $x\rightarrow -x$. 

In Fig.~\ref{fig4}(b) we use a crude approximation to show that the Ehrenfest Eqs.~\eqref{eq:xt}-\eqref{eq:sigma} would lead to skipping orbits at the edges for both trivial and nontrivial topological regimes. This would yield the absurd conclusion that one should expect edge states in both trivial and nontrivial topological regimes. These results are misleading. Here we correctly consider specular reflections at the edges by flipping the velocity sign when there is a collision, i.e., $k_y^0 \rightarrow -k_y^0$ and $\mean{\sigma_y} \rightarrow -\mean{\sigma_y}$ in Eq.~\eqref{eq:yt}, which is consistent with the pseudospin texture of the bulk band structure. However, the Ehrenfest equations only return a closed set of equations if one considers the plane wave limit, such that $\bm{k}$ is a good quantum number. In the opposite limit, for narrow wave packets, significant deviations are expected \cite{Schliemann2007SideJump}. This means that to use the Ehrenfest equations above, one must assume that they describe the center of motion of a broad packet. Moreover, for reasonable values of $M$ ($= \pm 10$~meV), the finite side-jump is only $|y(\infty)| \approx 20$~nm, which is much smaller than the required packet broadening ($\sim 1$~$\mu$m). Therefore the trajectories in Fig.~\ref{fig4}(b) are not compatible with the approximations that validate Eqs.~\eqref{eq:xt}-\eqref{eq:sigma}, hence the skipping orbits shown in Fig.~\ref{fig4}(b) are not accurate.

Nonetheless, the Ehrenfest Eqs.~\eqref{eq:xt}-\eqref{eq:sigma} can be applied for a finite ribbon of width $W$ if one considers the initial packet to be a broad enough Gaussian packet. This requires $W$ to be even larger, so that the ribbon can accommodate the initial packet. For the typical set of parameters, this requires an initial Gaussian broadening $\Gamma > 1$~$\mu$m, which also guarantees that the packet broadening is nearly constant in time.  The resulting motion is to to be seen as qualitative, since it does not consider the broadening explicitly. 

Let us consider $W = 4$~$\mu$m. Since the side-jump is in the nanometer range, an initial packet with $k_y^0 = 0$  would never reach the edges at $y = \pm W/2$. Hence  a finite $k_y^0 \neq 0$ is necessary to lead to a collision of the packet with the edge. The resulting dynamics are shown in Fig.~\ref{fig4}(c). At the edges $y = \pm W/2$, we consider the specular reflection described above. For any set of realistic parameters, we see the trajectories bouncing at the edges with no signature of skipping orbits. The difference between the trajectories in Fig.~\ref{fig4}(c) arise from the broken electron-hole symmetry of $H$, while for $D=0$ they become identical. These results agree with the full quantum dynamics of a Gaussian wave packet shown in Fig.~\ref{fig5}, which we discuss next.

\section{Wave packet dynamics}
\label{sec:wavepacket}

The Houston states provide a clear interpretation of the couplings $\mathcal{A}_{k_x(t)}^{n,n'}$ that lead to the LZT, while the zitterbewegung, calculated via the Ehrenfest equations, gives us insight about the dynamics of a wave packet. To complement these results, we now solve the time-dependent Schr\"odinger equation numerically to observe the dynamics of a wave packet as it collides with the edge of the system. We are interested in wide ribbons in the micrometer range, so here we consider the memory-efficient split-operator method \cite{foot1}, which is based on the Suzuki-Trotter expansion \cite{trotter1959product, suzuki1976generalized}. The approximate time-evolution operator is given by

\begin{equation}
 U(t+\tau, t) \approx e^{-i V_y \frac{\tau}{2\hbar}} e^{-i T_k(t) \frac{\tau}{\hbar}} e^{-i V_y \frac{\tau}{2\hbar}} + \mathcal{O}(\tau^3),
\end{equation}
where $V_y = M(y)\sigma_z$ is the single $y$-dependent term of $H_t$ [see Eq.~\eqref{eq:mass}], while $T_k(t)$ contains all $k$-dependent contributions from Eq.~\eqref{eq:BHZ}. For the initial state, Eq.~\eqref{eq:psi} now becomes a Gaussian wave packet with broadening $\Gamma$, and initial momentum $(k_x^0, k_y^0)$,  which reads

\begin{equation}
 \psi(y,0) = e^{i k_y^0 y}\dfrac{\exp(-\frac{y^2}{4\Gamma^2})}{(2\pi\Gamma^2)^{1/4}} \xi_\sigma,
\end{equation}
where $\xi_\sigma$ is the vector representation of the initial spin $\bm{\sigma}^0$, e.g. $\xi_\Uparrow = \binom{1}{0}$, $\xi_\Downarrow = \binom{0}{1}$. As time flows, the packet broadens approximately as $\Gamma_t = \sqrt{\Gamma^2 + [2 (D\pm B) t/\Gamma\hbar]^2}$. As usual, an initially narrow packet broadens quickly, while for initial $\Gamma \gg [2(D\pm B)t_\text{f}/\hbar]^{1/2}$ it remains $\Gamma_t \approx \Gamma$ within $0 < t < t_\text{f} \sim 500$~ps. For our typical set of parameters this condition requires $\Gamma \geq 1$~$\mu$m. Hereafter we consider $\Gamma = 1$~$\mu$m.

\begin{figure}[ht!]
  \centering
  \includegraphics[width=\columnwidth,keepaspectratio=true]{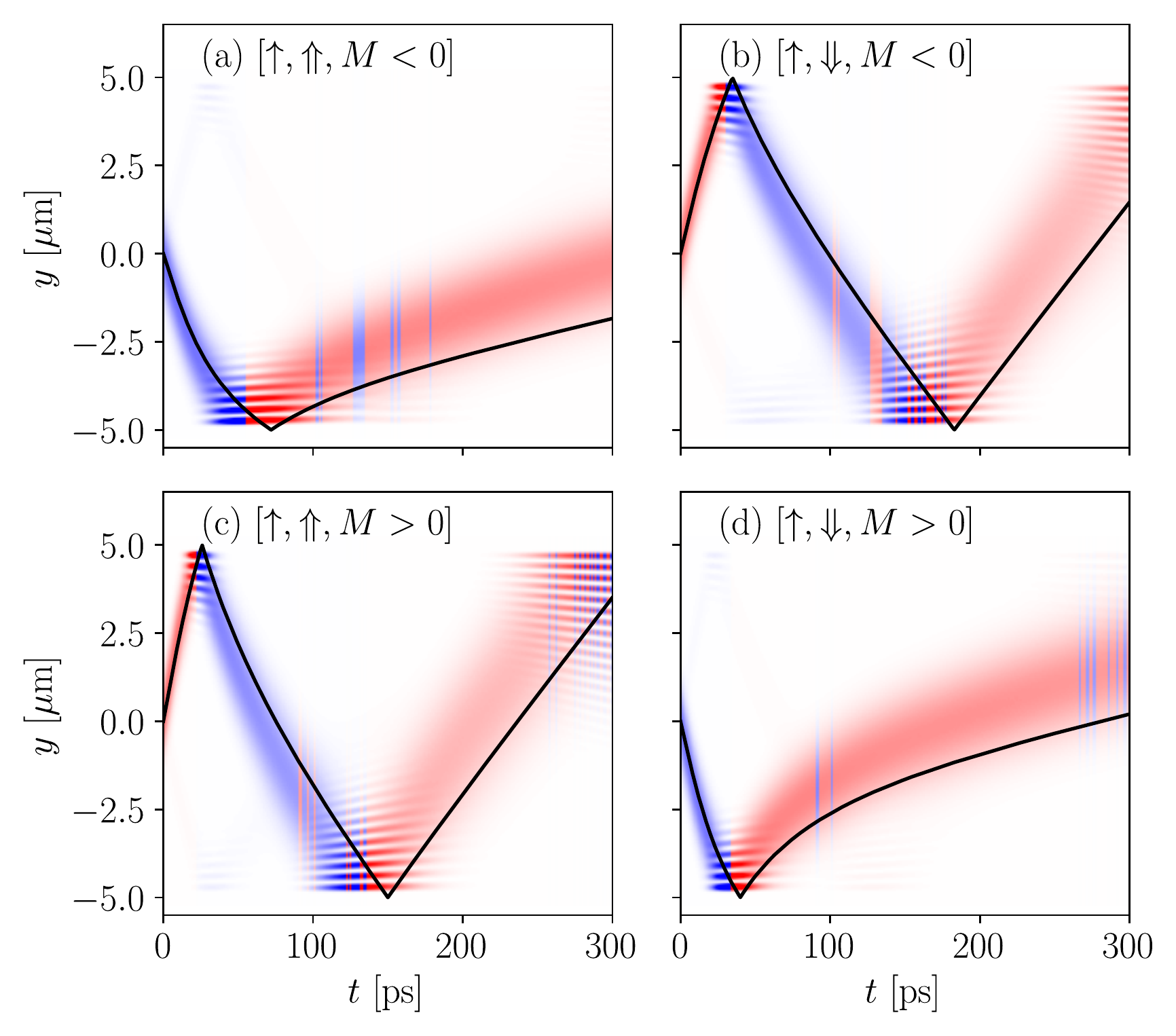}
  \caption{Time evolution of the density $|\psi(y,t)|^2$ and its corresponding Ehrenfest trajectory $\mean{y}(t)$ (black line) for a $W = 10$~$\mu$m ribbon [other parameters are equivalent to those in Fig.~\ref{fig4}(c)]. The spin is $s = \uparrow$, and the initial pseudo-spin $\sigma_z = \{\Uparrow, \Downarrow\}$ is indicated in each panel. The top (bottom) panels show the dynamics in the nontrivial (trivial) regime, $M = -10$~meV ($M = 10$~meV). In all cases the packets bounce at the edges $y = \pm W/2$. The blue and red colors indicate the sign of $\mean{\sigma_y}$ (time averaged over a few periods around each $t$). A small numerical noise is seen in the color code due to the fast oscillation of $\mean{\sigma_y}.$
  }
  \label{fig5}
\end{figure}

For $k_y^0 = 0$ the transverse motion is limited to the $\sim 20$~nm side-jump, which is much smaller than $\Gamma \sim 1$~$\mu$m, making it difficult to visualize the overall motion. Nonetheless, in this case the center of motion of the wave-packet matches the zitterbewegung trajectories, Figs.~\ref{fig1}(c) and~\ref{fig4}(a), obtained with the Ehrenfest equations. Due to the short side-jump, this bulk dynamics does not reach the edges.

For $k_y^0 \neq 0$ the results are shown in Fig.~\ref{fig5}. One immediately sees that the packet bounces off the borders unaffected by the presence of the edge states. Essentially there is no qualitative difference between the dynamics in the trivial ($M>0$) and nontrivial ($M<0$) topological regimes. Moreover, in all cases, the packet evolution agrees well with the Ehrenfest dynamics for equivalent initial conditions (black lines in Fig.~\ref{fig5}). The good agreement with the Ehrenfest trajectories emphasizes that edge states play no role in the dynamics of a packet initially set in bulk. The color code of the packet densities in Fig.~\ref{fig5} indicates the sign of $\mean{\sigma_y}$ obtained from the Schrödinger time-evolution. As the packet bounces off the borders, $\mean{\sigma_y} \rightarrow -\mean{\sigma_y}$ and $k_y \rightarrow -k_y$, thus justifying the specular reflection introduced in the Ehrenfest dynamics above.

\newpage 

\section{Conclusions}
\label{sec:conclusion}

We have shown that topological edge states effectively do not couple to bulk states via LZT. This conclusion arises exactly from the Houston function approach for the time-evolution that relates the LZT to the Berry connection matrix element $\mathcal{A}_{k_x}^{n,n'} \propto W^{-p}$, with $p=3/2~(1)$ for small (large) $k_x$. Numerical evaluation of $\mathcal{A}_{k_x}^{n,n'}$ confirms these scalings. Additionally, the zitterbewegung and numerical wave packet time-evolution were developed to further investigate the dynamics. These show packets bouncing off the edges in both trivial and nontrivial topological regimes. More interestingly, the zitterbewegung dynamics show that all possible initial conditions lead to a finite ballistic side-jump. Overall, these results contrast those from Ref.~\cite{Shi2013zitter}, where the zitterbewegung was introduced as a semiclassical picture for the topological helical edge states that yield the edge magnetization of the QSHE. Instead, for narrow ribbons the zitterbewegung can be associated with a ballistic SHE \cite{Xu2006zitterSHE}.

In a diffusive regime, we expect this dynamics to be  consistent with the Rashba-Edelstein effect \cite{Schliemann2007SideJump,Vignale2016Edelstein}, which might lead to spin accumulation at the edges (i.e., SHE) in both trivial and nontrivial topological regimes. This yields an interesting scenario, where there could be an interplay between the QSHE in SHE, depending on the chemical potential and how the electrons are injected into the sample. It is important to emphasize that here we have considered only the edge-bulk coupling via LZT, while other couplings could play significant role \cite{Stano2017NuclearSpin}. The full diffusive dynamics in a ribbon geometry with bulk and edge states considered on an equal footing remains both unexplored and challenging.

\acknowledgments The authors acknowledge support from CAPES, CPNq, FAPEMIG, FAPESP, and NAP Q-NANO from PRP/USP.

\bibliography{zitter}

\begin{thebibliography}{38}
\bibitem [{foot1()}]{foot1}%
  \BibitemOpen
  \bibinfo {note} {$V_y$ is diagonal in the real ($y$) space, while $T_k(t)$ is
  diagonal in the reciprocal $(k_y)$ space. To alternate between $y$ and $k_y$
  spaces we use FFT algorithms and apply each term of the time-evolution
  operator either in $y$- or $k_y$-space, hence minimizing the memory use.
  Additionally, since $T(t)$ is time-dependent, we consider a sufficiently
  short time-step $\tau $ to assure numerical convergence.}\BibitemShut {Stop}%
\end{thebibliography}

\end{document}